\documentclass{article}
\usepackage{spconfa4,amsmath,graphicx}

\usepackage{comment,multirow,makecell}
\usepackage{amssymb}
\usepackage{booktabs}
\usepackage{url} 
\usepackage[table,xcdraw]{xcolor}
\definecolor{darkgray}{rgb}{0.3, 0.3, 0.3}
\definecolor{a}{rgb}{0.92, 0.97, 1}  
\definecolor{b}{rgb}{0.8, 0.85, 0.9}  
\usepackage{enumitem}
\title{ImmersiveFlow: Stereo-to-7.1.4 Spatial Audio Generation with Flow Matching}
\name{
Zining Liang$^{1}$,
Runbang Wang$^{1,2}$,
Yujia Xiao$^{1}$,
Xuzhou Ye$^{3\dagger\thanks{$\dagger$ Corresponding authors.}}$,
Qiuqiang Kong$^{1\dagger}$
}
\address{
$^{1}$The Chinese University of Hong Kong, Hong Kong SAR, China\\
$^{2}$Nanjing University, Nanjing, China\\
$^{3}$ByteDance\\
}

\begin{document}
\ninept
\maketitle

\begingroup
\renewcommand{\thefootnote}{}
\footnotetext{\phantom{$\dagger$ }Demo: \url{https://immersiveflow.github.io}. Code: \url{https://github.com/violet-liang/ImmersiveFlow}.}
\endgroup

\begin{abstract}
Immersive spatial audio has become increasingly critical for applications ranging from AR/VR to home entertainment and automotive sound systems. However, existing generative methods remain constrained to low-dimensional formats such as binaural audio and First-Order Ambisonics (FOA). Binaural rendering is inherently limited to headphone playback, while FOA suffers from spatial aliasing and insufficient resolution for high-frequency.
To overcome these limitations, we introduce ImmersiveFlow, the first end-to-end generative framework that directly synthesizes discrete 7.1.4 format spatial audio from stereo input. ImmersiveFlow leverages Flow Matching to learn trajectories from stereo inputs to multichannel spatial features within a pretrained VAE latent space. At inference, the Flow Matching model predicted latent features are decoded by the VAE and converted into the final 7.1.4 waveform. 
Comprehensive objective and subjective evaluations demonstrate that our method produces perceptually rich sound fields and enhanced spatial impression, significantly outperforming traditional upmixing techniques. 
\end{abstract}
\begin{keywords}
Spatial audio generation, flow matching, immersive sound, generative models
\end{keywords}
\section{Introduction}
\label{sec:intro}

\begin{figure}[t]
    \centering
    \includegraphics[width=0.4\textwidth,height=0.52\linewidth]{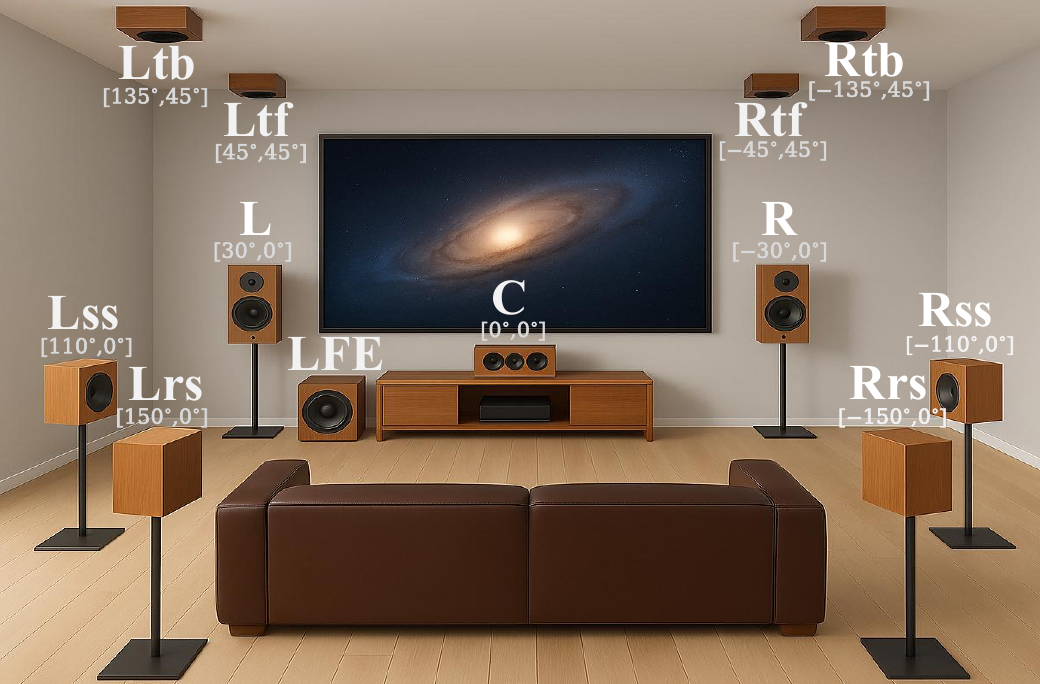}
    \caption{Illustration of the 7.1.4 loudspeaker configuration with positions, following ITU standard~\cite{ITU-RBS2051-3}. Abbreviations: L/R (Left/Right), C (Center), LFE (Subwoofer), Lss/Rss (Side Surround), Lrs/Rrs (Rear Surround), Ltf/Rtf (Top Front), Ltb/Rtb (Top Back). The positions shown are [azimuth, elevation].}
    \vspace{-0.6cm}
    \label{fig:714_speaker_layout}
\end{figure}

The demand for immersive spatial audio is rapidly growing, with two primary playback formats emerging. In personal applications such as AR/VR, spatial cues are rendered as binaural audio through headphones~\cite{zhu2025asaudio}. In shared environments such as home entertainment systems and automotive cabins, three-dimensional sound fields are reproduced using multichannel loudspeaker arrays, for example, the widely adopted 7.1.4 layouts~\cite{ITU-RBS2051-3} as illustrated in Figure ~\ref{fig:714_speaker_layout}. 
With recent advances in generative audio models~\cite{liu2024audioldm}, the task of spatial audio generation has gained increasing attention.
While existing studies primarily focus on sparse-channel formats such as binaural audio or First-Order Ambisonics (FOA), the direct synthesis of discrete, high-count immersive sound layouts like 7.1.4 remains both challenging and underexplored.

Recent research in generative spatial audio has primarily advanced along two streams. The first is dedicated to the synthesis of binaural audio for headphone listeners, which simulates spatial perception by modeling the Head-Related Transfer Functions (HRTFs). 
Early approaches adopt end-to-end frameworks that estimate binaural signals from monaural input, guided by spatial priors such as source positions inferred from visual content~\cite{lin2021exploiting,xu2021visually}. 
More recently, Both Ears Wide Open~\cite{sun2024both} introduces a diffusion-based method conditioned on text and image input to synthesize binaural audio, allowing complex spatial scenes with source movements.
The second stream focuses on the synthesis of First-Order Ambisonics (FOA), a four-channel spatial audio encoding format that can be decoded for multichannel playback. Diffusion-based models have been proposed to synthesize FOA using textual and parametric spatial descriptors conditions, such as sound category, source direction~\cite{heydari2024immersediffusion,kushwaha2024diff}. OmniAudio~\cite{liu2025omniaudio} further advances by conditioning on stereo audio and 360-degree visual inputs, incorporating scene semantics and spatial context from panoramic video.

Despite recent progress, binaural audio remains restricted to headphone playback and exhibits limited applicability in speaker-driven environments, as the spatial cues are encoded through HRTFs~\cite{braren2020high} that are only valid under headphone reproduction and cannot be accurately reproduced through multi-speaker setups. On the other hand, while FOA is versatile, it suffers from limited spatial resolution due to spatial aliasing, where high-frequency components roll off and the accuracy of sound field representation degrades~\cite{zotter2019ambisonics}. Existing solutions for multichannel output from low-dimension channel inputs often rely on traditional signal processing techniques, such as fixed-delay panning and matrix-based upmixing~\cite{wavdsp_upmix,nugen_halo}. These methods are computationally efficient yet lack perceptual realism~\cite{zhu2025asaudio}. Recent developments in flow-based generative modeling show more stable training and efficient inference compared to diffusion models~\cite{liu2025omniaudio, yuan2025flowsep, yun2025flowhigh}. However, its potential for high-channel spatial audio generation has not yet been explored.

Building on these insights, we propose \textit{ImmersiveFlow}, a novel generative model that produces 7.1.4-format audio from stereo. The model is trained using Conditional Flow Matching (CFM), employing a transformer-based vector field estimator to model high-channel spatial audio distributions. The generated output can be directly applied on loudspeaker arrays or binauralized for headphone playback. 
Our main contributions are summarized as follows: 
First, we introduce the novel task of multichannel spatial audio generation, and propose the first end-to-end stereo-to-7.1.4 model, overcoming the fundamental limitations of traditional spatial audio generation methods. 
Second, We pioneer Conditional Flow Matching (CFM) on VAE latents for high-channel spatial audio generation, employing a data-space objective to provide more stable and robust training.
Third, we introduce Spatial Energy Correlation (SEC), a novel objective metric for assessing the spatial fidelity of multichannel audio. Extensive experiments using both objective and subjective metrics demonstrate that our method achieves state-of-the-art performance, generating perceptually rich and immersive sound fields.


\begin{figure}[t]
  \centering
  \includegraphics[scale=0.8]{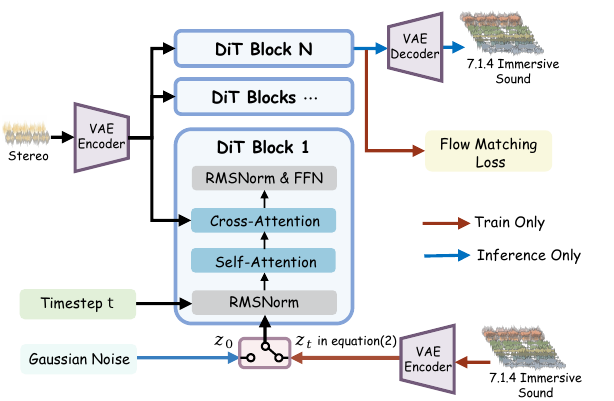}
  \vspace{-0.3cm}
  \caption{Overview of the ImmersiveFlow architecture. During training, both stereo and 7.1.4 immersive audio are encoded into latent using a pretrained VAE encoder. The model learns to map stereo latents to immersive sound latents using Flow Matching. During inference, only the stereo is input to predict immersive sound latents, which are then decoded to 7.1.4 audio using the VAE decoder.
  }
  \label{network}
  \vspace{-0.5cm}
\end{figure}

\section{METHOD}
\label{sec:METHOD}

We propose ImmersiveFlow, a generative framework based on Conditional Flow Matching for stereo-to-7.1.4 spatial audio. The model is composed of two key components: a pretrained Variational Autoencoder (VAE), which encodes the individual channels of both stereo and 7.1.4 audio into latent representations, and Conditional Flow Matching, which learns the mapping from stereo latent representations to immersive sound latents. The overall architecture is illustrated in Figure~\ref{network}.

\subsection{VAE Encoder and Decoder}
\label{sec: VAE Encoder and decoder}
As shown in Figure~\ref{fig:714_speaker_layout}, a 7.1.4 configuration involves 12 distinct speakers, which makes directly modeling its raw waveform computationally challenging.
While traditional methods typically convert audio into mel-spectrograms before compression, recent studies have shown that encoding directly at the waveform level using VAE lead to better reconstruction quality~\cite{evans2024fast}.
Therefore, we employ a pretrained VAE~\cite{lei2025levo} to transform input audio into a compact latent representation. 
Specifically, each channel is independently processed by the VAE encoder to obtain a latent $\mathbf{z} \in \mathbb{R}^{D \times T'}$, where $D$ denotes the latent dimension and $T'$ represents the number of frames. 
The latent representations from the respective channels are then concatenated. 
For immersive sound inputs, this leads to $\mathbf{z}_1 \in \mathbb{R}^{C \times D \times T'}$, where $C = 12$ corresponds to the 7.1.4 channel layout, serving as the target for the generative model. For stereo inputs, the resulting latent representation is $\mathbf{z}_{\text{cond}} \in \mathbb{R}^{C \times D \times T'}$, where $C = 2$ indicating the left and right channels, which act as the conditioning input to the generative model.

\subsection{Flow Matching for Immersive Sound Generation}
\label{sec: Flow Matching for immersive sound generation}
We adopt Conditional Flow Matching as our generative backbone, which models a continuous transformation from a prior distribution to the target data distribution. CFM defines a probability flow governed by a time-dependent velocity field. The trajectory of a sample $\mathbf{z}_t$ over normalized time $t \in [0, 1]$ is described by the following ordinary differential equation (ODE):
\begin{equation}
\frac{d \mathbf{z}_t}{dt} = \mathbf{v}(\mathbf{z}_t, t),
\end{equation}
where $\mathbf{v}(\mathbf{z}_t, t)$ is the velocity field.

During training, we define a linear interpolation path between a sampled noise vector $\mathbf{z}_0 \sim \mathcal{N}(0, I)$ and the target latent $\mathbf{z}_1$ as:
\begin{equation}
\mathbf{z}_t = (1 - t) \mathbf{z}_0 + t \mathbf{z}_1, \quad t \in [0, 1]
\label{1}
\end{equation}
The target velocity field $\mathbf{u}(\mathbf{z}_t | \mathbf{z}_0, \mathbf{z}_1)$ at any intermediate point $\mathbf{z}_t$ along the linear path is defined as:
\begin{equation}
\mathbf{u}(\mathbf{z}_t| \mathbf{z}_0, \mathbf{z}_1)=\mathbf{z}_1 - \mathbf{z}_0.
\label{2}
\end{equation}

We parameterize the velocity field using a transformer-based network following the Denoising Diffusion Transformer (DiT) design~\cite{peebles2023scalable}, denoted as $\mathbf{v}_\theta(\mathbf{z}_t, t, \mathbf{z}_{\text{cond}})$, where $\theta$ represents the model parameters and $\mathbf{z}_{\text{cond}}$ is the stereo latent input. The DiT model comprises $N$ stacked transformer blocks, each consisting of self-attention, cross-attention, and feedforward layers, with time-step conditioning and latent normalization applied throughout. The stereo latent $\mathbf{z}_{\text{cond}}$ is conditioned through feature-wise linear modulation (FiLM)~\cite{perez2018film} to guide the transformation toward the target audio representation.


Recent studies~\cite{yao2025flow2gan,li2025back} suggest that shifting the optimization target from the velocity field to the data space can leverage the structural priors of clean signals, providing a more consistent and robust training signal for audio synthesis. In this work, we employ a data-space optimization objective for ImmersiveFlow.
Specifically, the prediction of clean data $\mathbf{x}_\theta$ can be analytically derived from the predicted velocity $\mathbf{v}_\theta$ through the linear interpolation path in Eq.(\ref{1}):
\begin{equation}
\mathbf{x}_\theta = (1 - t) \mathbf{v}_\theta + \mathbf{z}_t.
\label{x}
\end{equation}
The resulting training objective is defined as the mean squared error between the predicted and target clean latents:
\begin{equation}
\mathcal{L}_{\text{FM}} = \mathbb{E}_{t, \mathbf{z}_0, \mathbf{z}_1} || \mathbf{x}_\theta(\mathbf{z}_t, t, \mathbf{z}_{\text{cond}}) - \mathbf{x_1} ||_2^2,
\label{x-loss}
\end{equation}
where $\mathbf{x}_1$ denotes the target clean latent, which corresponds to the terminal point $\mathbf{z}_1$ in Eq. (\ref{1}).

\subsection{Solving ODE for Inference}
\label{sec: Solving the Ordinary Differential Equation for Inference}
Inference is carried out by solving an Ordinary Differential Equation (ODE) specified by the learned velocity field. Starting from an initial vector $\mathbf{z}_0 \sim \mathcal{N}(\mathbf{0}, \mathbf{I})$, the trajectory is computed conditioned on the stereo latent input $\mathbf{z}_{\text{cond}}$:
\begin{equation}
\mathbf{z}_{t+\Delta t} = \mathbf{z}_t + \mathbf{v}_\theta(\mathbf{z}_t, t, \mathbf{z}_{\text{cond}}) \Delta t,
\label{eq:euler}
\end{equation}
where $\Delta t$ denotes the integration step size. Here, we use the Dormand–Prince solver to compute the final latent at $t = 1$.

Unlike diffusion-based models that rely on hundreds of iterative denoising steps, this formulation enables fast inference through direct velocity regression, substantially reducing computational cost while preserving perceptual quality.
Once the flow matching module predicts the immersive sound latent, the VAE decoder restores it into the 7.1.4 audio waveform.


\section{EVALUATION}
\label{sec:EVALUATION}
\subsection{Dataset}
\label{sec:Dataset}
Due to the scarcity of publicly available high-quality 7.1.4 format audio, we curated a collection of 500 professionally mixed and mastered songs.
Each track is approximately three minutes in duration and sampled at 48 kHz. The recordings encompass a wide variety of musical elements, including lead and backing vocals, drums, guitars, and diverse electronic effects. 
For training, we generate paired samples by downmixing each 7.1.4-channel audio track to its stereo equivalent using a fixed matrix based on the AC-3 standard~\cite{standard2012digital}. 
During preprocessing, we extract ten 10-second segments with randomized offsets from each track. These segments are then projected into a latent space using a pre-trained VAE, with the corresponding latent representations pre-extracted for subsequent training.
The dataset was partitioned by song to ensure no overlap between sets, with 90\% allocated for training and 10\% reserved for testing. 

\begin{table*}[!htbp]
\centering
\caption{Performance comparison between our proposed models and the baselines across each 7.1.4 channel on the test sets. 
We report two perceptual audio quality metrics, Virtual Speech Quality Objective Listener (ViSQOL) and PAM, 
as well as two generative metrics, Fréchet Audio Distance (FAD), and MAUVE Audio Distance (MAD). 
 }
\label{tab:obj_metrics}
\resizebox{\linewidth}{!}{%
\newcommand{\cbox}[2]{\cellcolor{#1}#2}
\begin{tabular}{llcccccccccccc}
\toprule
Metric & Model & L & R & C & LFE & Lss & Rss & Lrs & Rrs & Ltf & Rtf & Ltb & Rtb \\
\midrule
\multirow{4}{*}{\makecell{$\text{ViSQOL} \uparrow$}} 

& WavDSP Upmix~\cite{wavdsp_upmix}   
& 3.716 & 3.686 & 4.013 & 3.917 
& 3.270 & 3.292 & 4.039 & 4.048 
& 3.665 & 3.688 & 3.807 & 3.829 \\

& Halo Upmix~\cite{nugen_halo}      
& 4.071 & 4.068 & 4.011 & 4.105 
& \textbf{3.291} & \textbf{3.305} & 4.098 & 4.092 
& 3.627 & 3.613 & 3.665 & 3.666 \\

& \cbox{a}{ImmersiveFlow} 
& \cbox{a}{3.927} & \cbox{a}{3.923} & \cbox{a}{3.976} & \cbox{a}{4.017} 
& \cbox{a}{3.240} & \cbox{a}{3.251} & \cbox{a}{4.167} & \cbox{a}{\textbf{4.174}} 
& \cbox{a}{3.869} & \cbox{a}{3.874} & \cbox{a}{3.913} & \cbox{a}{3.915} \\

& \cbox{b}{ImmersiveFlow-pair}   
& \cbox{b}{\textbf{4.139}} & \cbox{b}{\textbf{4.105}} & \cbox{b}{\textbf{4.051}} & \cbox{b}{\textbf{4.297}} 
& \cbox{b}{3.242} & \cbox{b}{3.269} & \cbox{b}{\textbf{4.176}} & \cbox{b}{4.161} 
& \cbox{b}{\textbf{4.126}} & \cbox{b}{\textbf{4.140}} & \cbox{b}{\textbf{4.133}} & \cbox{b}{\textbf{4.145}} \\


\midrule

\multirow{5}{*}{\makecell{$\text{PAM} \uparrow$}}
& \textcolor{darkgray}{\textit{GT}} & \textcolor{darkgray}{\textit{0.869}} & \textcolor{darkgray}{\textit{0.880}} & \textcolor{darkgray}{\textit{0.631}} & \textcolor{darkgray}{\textit{0.558}} 
& \textcolor{darkgray}{\textit{0.793}} & \textcolor{darkgray}{\textit{0.815}} & \textcolor{darkgray}{\textit{0.728}} & \textcolor{darkgray}{\textit{0.757}} 
& \textcolor{darkgray}{\textit{0.765}} & \textcolor{darkgray}{\textit{0.751}} & \textcolor{darkgray}{\textit{0.788}} & \textcolor{darkgray}{\textit{0.769}} \\

& WavDSP Upmix~\cite{wavdsp_upmix} 
& 0.802 & 0.804  & 0.730 & \textbf{0.537} 
& 0.674 & 0.696 & \textbf{0.722} & \textbf{0.716} 
& 0.727 & 0.728 & 0.745 & 0.742\\

& Halo Upmix~\cite{nugen_halo} 
& \textbf{0.811} & \textbf{0.838}  & \textbf{0.815} & 0.523 
& \textbf{0.763} & \textbf{0.765} & 0.700 & 0.703 
& 0.354 & 0.354 & 0.354 & 0.354\\

& \cbox{a}{ImmersiveFlow} & \cbox{a}{0.747} & \cbox{a}{0.746}  & \cbox{a}{0.687} & \cbox{a}{0.531} 
& \cbox{a}{0.646} & \cbox{a}{0.645} & \cbox{a}{0.680} & \cbox{a}{0.681} 
& \cbox{a}{0.682} & \cbox{a}{0.684} & \cbox{a}{0.673} & \cbox{a}{0.679}\\

& \cbox{b}{ImmersiveFlow-pair} & \cbox{b}{0.803} & \cbox{b}{0.804}  & \cbox{b}{0.656} & \cbox{b}{0.536}
& \cbox{b}{0.726} & \cbox{b}{0.729} & \cbox{b}{0.701} & \cbox{b}{0.705}
& \cbox{b}{\textbf{0.738}} & \cbox{b}{\textbf{0.732}} & \cbox{b}{\textbf{0.703}} & \cbox{b}{\textbf{0.700}}\\



\midrule

\multirow{4}{*}{\makecell{${\text{FAD}}_{\textcolor{gray}{\text{\tiny{CLAP}}}} \downarrow$}}

& WavDSP Upmix~\cite{wavdsp_upmix} & 0.119 & 0.126  & 0.816 & 1.026 & 0.239 & 0.237 & 0.639 & 0.620 & 0.470 & 0.458 & 0.535 & 0.532\\

& Halo Upmix~\cite{nugen_halo}    & 0.042 & 0.043  & 0.511 & 0.374 & 0.200 & 0.210 & 0.367 & 0.385 & 1.143 & 1.140 & 1.056 & 1.044\\

& \cbox{a}{ImmersiveFlow} & \cbox{a}{0.034} & \cbox{a}{0.035}  & \cbox{a}{0.131}  & \cbox{a}{0.143} 
& \cbox{a}{\textbf{0.111}} & \cbox{a}{\textbf{0.113}} & \cbox{a}{0.168} & \cbox{a}{0.172} 
& \cbox{a}{0.163} & \cbox{a}{0.165} & \cbox{a}{0.198} & \cbox{a}{0.188}\\

& \cbox{b}{ImmersiveFlow-pair} & \cbox{b}{\textbf{0.018}}  & \cbox{b}{\textbf{0.016}} & \cbox{b}{\textbf{0.086}} & \cbox{b}{\textbf{0.056}}
&\cbox{b}{ 0.131} & \cbox{b}{0.126} & \cbox{b}{\textbf{0.076}} & \cbox{b}{\textbf{0.189}}
& \cbox{b}{\textbf{0.111}} & \cbox{b}{\textbf{0.113}} & \cbox{b}{\textbf{0.127}} & \cbox{b}{\textbf{0.128}}\\



\midrule

\multirow{4}{*}{\makecell{${\text{MAD}}_{\textcolor{gray}{\text{\tiny{MERT}}}} \downarrow$}} 
& WavDSP Upmix~\cite{wavdsp_upmix} & 1.609 & 2.316 & 4.821 & 5.504 & 4.787 & 4.866 & 3.156 & 2.345 & 4.113 & 4.891 & 4.793 & 4.906 \\

& Halo Upmix~\cite{nugen_halo}    & 0.029 & 0.031 & 3.922 & 5.504 & 2.335 & 2.988 & 1.307 & 1.059 & 5.504 & 5.504 & 5.504 & 5.504 \\

& \cbox{a}{ImmersiveFlow} & \cbox{a}{0.026} & \cbox{a}{0.012} & \cbox{a}{1.598} & \cbox{a}{0.841} 
& \cbox{a}{\textbf{0.878}} & \cbox{a}{\textbf{0.705}} & \cbox{a}{1.166} & \cbox{a}{1.035}
& \cbox{a}{0.984} & \cbox{a}{0.650} & \cbox{a}{1.059} & \cbox{a}{0.938} \\

& \cbox{b}{ImmersiveFlow-pair} & \cbox{b}{\textbf{0.012}} & \cbox{b}{\textbf{0.007}} & \cbox{b}{\textbf{0.490}} &\cbox{b}{\textbf{0.798}}
& \cbox{b}{1.283} & \cbox{b}{1.123} & \cbox{b}{\textbf{1.069}} & \cbox{b}{\textbf{1.012}}
& \cbox{b}{\textbf{0.413}} & \cbox{b}{\textbf{0.472}} & \cbox{b}{\textbf{0.991}}& \cbox{b}{\textbf{0.885}} \\



\bottomrule
\end{tabular}%
}
\end{table*}

\subsection{Evaluation Metrics}
\label{sec:Evaluation Metrics}
We assess model performance in terms of both
objective metrics and subjective evaluation.

\noindent 1) \textit{Objective Evaluation}:
Conventional sample-level metrics such as PESQ, and SDR rely on strict waveform alignment and are unsuitable for evaluating generative models~\cite{yuan2025flowsep, wang2025flowse}.
Therefore, we adopt three categories of objective metrics: Perceptual quality metrics include Virtual Speech Quality Objective Listener (ViSQOL)~\cite{chinen2020visqolv3opensource} and PAM~\cite{deshmukh2024pam}. 
ViSQOL is a full-reference metric that estimates mean opinion scores (MOS), while PAM is a no-reference neural estimator that leverages audio-language models to predict quality. 
Manifold-based metrics include the widely used Fréchet Audio Distance (FAD)~\cite{kilgour2018fr} with CLAP~\cite{laionclap2023} embeddings, along with the complementary MAUVE Audio Distance (MAD)~\cite{huang2025aligning}, which is based on MERT~\cite{li2023mert} embeddings.
To evaluate spatial fidelity across scene depth, horizontal envelopment, and vertical envelopment~\cite{ITU-RBS1116-3}, we introduce \textbf{Spatial Energy Correlation (SEC) }. Multichannel signals are projected onto a dense spherical grid to compute spatial energy maps, which are then divided into three perceptual zones:: Front for scene depth ($\text{elevation} \leq 30^\circ$, $|\text{azimuth}| \leq 60^\circ$), Surround for horizontal envelopment ($\text{elevation} \leq 30^\circ$, $|\text{azimuth}| > 60^\circ$), and Height for vertical envelopment ($\text{elevation} > 30^\circ$). SEC is defined as the correlation coefficient between the synthesized and ground-truth energy maps for each zone.


\noindent 2) \textit{Subjective Evaluation}:
We conducted a subjective listening test to evaluate the overall audio quality and spatial impression of the generated 7.1.4 audio, using MOS as the primary evaluation metric. 
To enable broad-scale online testing, we used webMUSHRA~\cite{schoeffler2018webmushra} for headphone playback. Both the reference and generated 7.1.4 signals were binauralized using the KEMAR HRTFs from the SADIE II database~\cite{armstrong2018perceptual}.   
Following the speaker layout defined in Figure~\ref{fig:714_speaker_layout}, the binaural filters for each 7.1.4 channel were synthesized via VBAP~\cite{pulkki1997vbap} interpolation of the three nearest HRTFs, while the LFE channel was down-mixed by 6 dB to the binaural output. Loudness normalization was applied to all samples at -24.0 LUFS.
The evaluation set consisted of six 10-second clips, randomly selected from six songs in the test set. These clips were evenly divided into two scenarios: static scenes with fixed head positions, and dynamic scenes where a full clockwise head rotation was simulated by orbiting the source counter-clockwise.
Participants rated each method for both audio quality and spatial impression on a scale of $[0, 100]$. 

\subsection{Experiments Details}

We evaluate our proposed method against two commercial upmixers, \textbf{Halo Upmix}~\cite{nugen_halo} and \textbf{WavDSP UpMix}~\cite{wavdsp_upmix}. To thoroughly investigate the modeling strategy, we propose two variants of our model: \textbf{ImmersiveFlow} and \textbf{ImmersiveFlow-pair}. The standard ImmersiveFlow employs a joint configuration that models all 12 channels simultaneously. Alternatively, ImmersiveFlow-pair serves as a baseline that exploits spatial symmetry. It partitions the 7.1.4 layout into seven logical pairs: five symmetric pairs (L/R, Lss/Rss, Lrs/Rrs, Ltf/Rtf, and Ltb/Rtb) and two duplicated pseudo-pairs for the monophonic C and LFE channels. During inference, these pairs are synthesized in parallel under a shared stereo condition and subsequently aggregated into the standard 7.1.4 layout.

For our ImmersiveFlow model, we use a pretrained VAE from the SongGeneration~\cite{lei2025levo} checkpoint to encode both stereo and 7.1.4 audio into latent representations. The encoder operates at a temporal resolution of 25 frames per second, producing latent features with a dimensionality of $D = 64$. 
To map these latent features from stereo to 7.1.4, a DiT architecture is employed. The DiT backbone consists of $N = 12$ transformer blocks, each configured with 16 attention heads and a hidden dimension of 1024. In total, the model comprises approximately 281M trainable parameters.
During training, the VAE encoder and decoder are kept frozen, and only the DiT is optimized. 
ImmersiveFlow and ImmersiveFlow-pair share the same backbone architecture and training configurations, including a learning rate of $1 \times 10^{-4}$ and a batch size of 16 for 400k steps. 

\subsection{Results}
\label{sec:Results and Discussion}

\begin{figure}[h!]
  \centering
  \vspace{-0.3cm}
  \includegraphics[scale=0.45]{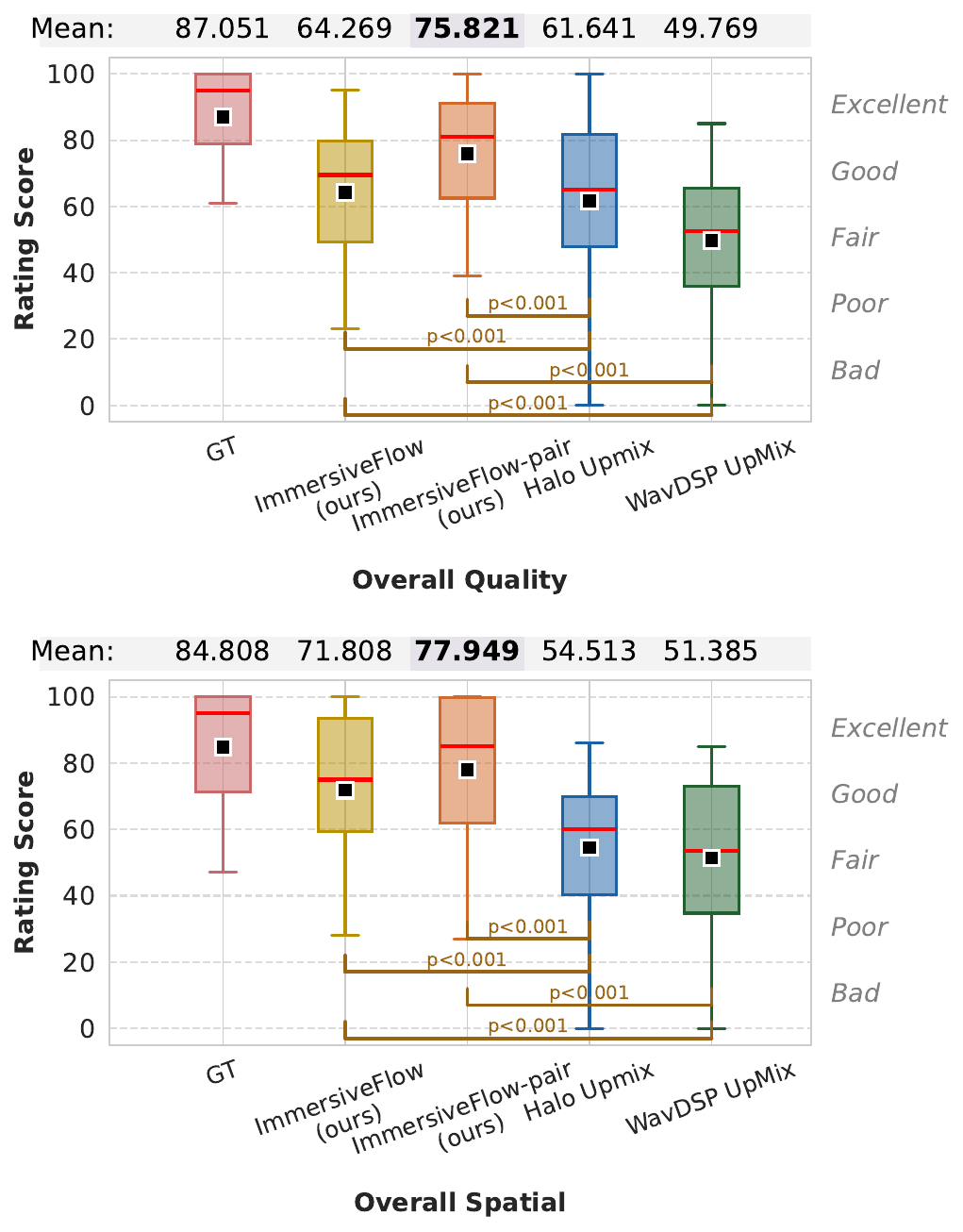}
  \vspace{-0.3cm}
  \caption{Subjective test score distributions for Overall Quality and Overall Spatial. Black squares indicate means, red lines indicate medians, and brackets denote pairwise significance.
  }
  \label{mushura}
  \vspace{-0.5cm}
\end{figure}

\begin{table}[h]
\vspace{-0.6cm}
\centering
\caption{Spatial Energy Correlation (SEC) comparison between our proposed models and the baselines.}
\label{tab:seds}
\begin{tabular}{llccc}
\toprule
Metric & Model                & Front   & Surround & Height  \\
\midrule
\multirow{4}{*}{SEC $\uparrow$}
       & Halo Upmix          & 0.9447  & 0.8633   & 0.8161  \\
       & WavDSP Upmix        & 0.8418  & 0.7606   & 0.6689  \\
       & ImmersiveFlow       & 0.9495  & 0.9134   & 0.8597  \\
       & ImmersiveFlow-pair  & \textbf{0.9695} & \textbf{0.9303} & \textbf{0.9429} \\
\bottomrule
\end{tabular}
\vspace{-0.35cm}
\end{table}

\noindent 1) \textit{Objective Evaluation}:
Table~\ref{tab:obj_metrics} and Table~\ref{tab:seds} summarizes the objective evaluation results of our proposed method, ImmersiveFlow compared to baselines on the test set, with results for each generated 7.1.4 channel. Overall, both proposed variants demonstrate a clear advantage over the baselines, validating the effectiveness of our joint and pairwise generation framework.
For perceptual quality, the base ImmersiveFlow establishes a strong performance by outperforming commercial tools in rear surround (Lrs/Rrs) and height channels (Ltf/Rtf/Ltb/Rtb), validating the effectiveness of our joint modeling framework in capturing the overarching spatial atmosphere.
ImmersiveFlow-pair further elevates performance, achieving the highest ViSQOL scores across nearly all channels. This suggests that the model has mastered the functional roles of 7.1.4 channels: main channels (L/R/C) preserve harmonic and vocal clarity, the LFE concentrates low-frequency energy, and the surround/top channels retain ambient effects for immersion.
While our models exhibit slight PAM degradations due to fine-grained detail loss during VAE compression, the pair version significantly narrows this gap. Notably, Halo Upmix yields PAM scores that exceed the GT on some channels, often through the incorporation of enhancement modules (e.g., reverbs, EQ filters) to improve perceived quality, whereas our models prioritize high-fidelity reconstruction.
For manifold-based metrics (FAD, MAD), both variants significantly outperform all baselines, validating the model’s ability to reconstruct accurate spatial energy distributions and maintain spatial fidelity. ImmersiveFlow-pair achieves the best results by using a pairwise strategy to capture inter-channel correlations, including intricate phase relationships and energy balance.
For spatial fidelity (SEC), our proposed models achieve the highest scores, producing an overall energy distribution across all spatial zones. Conversely, commercial upmixers show a clear performance degradation in surround and height regions, tending to collapse spatial energy toward the front speakers and thus limiting natural envelopment.

\noindent 2) \textit{Subjective Evaluation}: 
Figure \ref{mushura} illustrates the results of the subjective listening test involving 13 trained participants, where both ImmersiveFlow and ImmersiveFlow-pair outperform commercial upmixing techniques in terms of audio quality and spatial impression. ImmersiveFlow-pair achieves the highest mean scores, while commercial baselines struggle with spatial rendering. 

\section{CONCLUSION}
\label{sec:CONCLUSION}
In this work, we present ImmersiveFlow, the first end-to-end generative framework capable of directly synthesizing discrete 7.1.4 format spatial audio from stereo input. By leveraging Flow Matching within a pretrained VAE latent space, ImmersiveFlow effectively models multichannel spatial features and reconstructs perceptually coherent waveforms. Extensive objective and subjective evaluations confirm that our method generates rich and immersive sound fields, substantially outperforming conventional upmixing techniques. This study not only presents a new paradigm for high-channel spatial audio generation but also highlights the potential of flow-based generative modeling for broader immersive media applications.

\bibliographystyle{IEEEbib}
\bibliography{refs}

\end{document}